\documentclass[preprint,12pt,longbibliography]{elsarticle}
\usepackage{amssymb}
\usepackage{amsfonts}
\usepackage{amsthm}
\usepackage{amsmath}
\usepackage{newlfont}
\usepackage{stmaryrd}
\usepackage{mathrsfs}
\usepackage{euscript}
\usepackage{graphicx}
\usepackage{enumerate}
 \usepackage[final]{changes}
 
\usepackage{wrapfig}

\usepackage{amscd}
\usepackage{hyperref}

\theoremstyle{plain}

\theoremstyle{definition}

\theoremstyle{remark}

\newcommand{\opunit}{\text{1}\kern-0.22em\text{l}}

\DeclareMathAlphabet{\mathpzc}{OT1}{pzc}{m}{it}

\newcommand{\id}{\textrm{d}}

\usepackage{changes}

\title{Specific heat of a driven lattice gas}
\author{Pritha Dolai\corref{cor1}}
\author{ Christian Maes}
\address{Department of Physics and Astronomy --- KU Leuven, Belgium}
\cortext[cor1]{pritha.dolai@kuleuven.be}

\begin{document}
\begin{abstract}
Calorimetry for equilibrium systems aims to determine the available microscopic occupation and distribution of energy levels by measuring thermal response.  Nonequilibrium versions are expected to add information on the dynamical accessibility of those states. We perform calculations on a driven exclusion process on an array of particle stations, confirming that expectation.
That Markov model produces a fermionic nonequilibrium steady state where the specific heat is computed exactly by evaluating the heat fluxes that  are entirely due to a change in ambient temperature.  We observe a zero-temperature divergence (violation of the Third Law) when the Fermi energy and the kinetic barrier for loading and emptying become approximately equal. Finally, when the kinetic barrier is density-dependent, a stable low-temperature regime of negative specific heat appears, indicating an anti-correlation between the temperature--dependence of the stationary occupation and the excess heat. 
\end{abstract}
\date{\today}
\maketitle

\section{Introduction}
Measuring heat produced in physical processes or chemical reactions, or, more generally, measuring the absorbed heat in some relaxation process is at the heart of calorimetry and thermal physics.  It has a long and important history, especially when the pioneers of kinetic theory realized that there were important discrepancies between the predictions of the specific heat based on the classical equipartition theorem and the experimental data where the specific heat was seen to vary and to vanish at very low temperatures.  A fundamental breakthrough was the application to solids of Planck's quantization hypothesis, as suggested by Einstein at the 1911 Solvay conference.  From that moment, quantization of energy entered the theory of condensed matter as well, and the old problem of specific heats disappeared. Yet, and still today, there are new challenges for the theory of specific heats, and calorimetry will continue to play an important role in a great variety of domains ranging from climate science, over biophysics to black hole thermodynamics.\\

In  recent years, attempts have been made to include steady nonequilibrium processes, as they are relevant to biophysics, material science,  and far-from-equilibrium transport.  Again, both applied and fundamental issues arise, in particular, to understand how {\it functionalities} (not merely passive material properties) are measurable from nonequilibrium calorimetry. By functionalities we mean mostly the (working) presence of pumps and motors, realizing steady currents and active processes.  Nonequilibrium may refer to transient regimes (on the way to relaxation to equilibrium), or to (bulk and boundary) driven and to (individually) active particles in steady regimes. The latter include biophysical systems such as bacteria and molecular motors that are (mostly) chemically fueled, while driving refers to the application of nongradient forces to open systems with steady transport.  Yet, because of the absence of global thermodynamic potentials, the thermal properties of nonequilibrium systems remain theoretically less understood.   The main difficulty is that the heat released in system transitions may no longer teach us the same information about the occupation statistics and energy degeneracies as under thermal equilibrium. Outside equilibrium, there appears a divergence between the Clausius notion of entropy which is directly related to heat, and the Boltzmann notion of entropy, which is configurational and estimates microstate multiplicity.  This problem is general and of fundamental importance in evaluating entropies and heat capacities for systems as diverse as evaporating black holes and bacterial colonies.  When it really comes to generic nonequilibrium situations, away from special symmetries (such as time-reversal or supersymmetry) new approaches are needed to probe the thermal physics.\\
The present paper takes up that challenge for a specific many-body system.\\
 
The application of calorimetry to far-from-equilibrium systems is not only more problematic; it is revealing as well.  It is indeed reasonable to expect that their thermal response exposes kinetic (non-thermodynamic) information, especially at low temperatures. We wish therefore to understand how aspects of heat capacity directly connect with the statistical features of the nonequilibrium dynamics, not just with the static fluctuations as is the case in equilibrium.\\
Our model will illustrate those points in considerable detail as an exact solution is available.  As will be clearly seen and is relevant to driven systems more generally, \cite{dlg1,dlg2,dlg3}, the specific heat becomes a function of thermodynamic but also of kinetic, and driving parameters.\\

After the start of steady state thermodynamics, \cite{kom2,prig,oon}, nonequilibrium heat capacities have been introduced, \cite{eu,jir, jstat, ner,ner2,nernst2}. 
However, so far they were computed explicitly for (effectively) independent particles only. The present paper extends that to driven lattice gases where, subject to exclusion, particles hop between stations (dots) and are created or annihilated with density-dependent constraints.  The births and deaths arise from a coupling to a thermochemical bath at chemical potential $\mu$ and temperature $T$. The bath is also the reservoir for dissipating the Joule heat due to the particle-driving. 
We think here of modeling electron hopping between quantum dots, where the asymmetric exclusion process provides a Fermi Golden Rule approximation upon adding births and deaths at each dot.
We thus get a periodic array of two-level systems in which a particle current is maintained. 
In summary, this yields a semiclassical description of transport along an array of quantum dots (QD), \cite{cxz,km,mhcj,szafran}, fed by an electron bath; see Fig.\ref{twolevel}(a). We study its thermal response.\\

Coupling two-level systems in an array has obviously a great number of applications, {\it e.g.}, in heat engines \cite{jld} and quantum devices \cite{qdevice1,qdevice2}. 
We focus on the specific heat. 
Exact calculations of  heat capacities have also been performed in open quantum systems; see {\it e.g.} \cite{subasi} also containing a
discussion about the Third law of Thermodynamics.
Note however that in contrast with equilibrium, the heat capacity is not obtainable as temperature-derivative of a thermodynamic potential. That is because of the presence of irreversible work.  The idea is to look at the excess heat:  when the temperature of the heat bath is changed, the original nonequilibrium condition relaxes to the new nonequilibrium and produces ``extra'' heat on top of the stationary dissipated power.  In particular, changing the temperature $T \rightarrow T + \id T$, that excess heat $\delta Q = C(T)\,\id T$ defines the heat capacity $C(T)$ \cite{eu,jir,jstat,ner,ner2,active,exact}, in complete analogy with reversible transformations between equilibria, \cite{nernst2}.\\ 
As we consider a driven system in its steady state, the heat capacity $C(T ) = C(T, \zeta)$ will depend on the driving $\zeta$ as well, and only for $\zeta = 0$, when there is detailed balance, we recover the equilibrium heat capacity.
Besides giving an exact expression of $C(T)$, the present paper reports on two {\it unusual} properties with respect to equilibrium, visible in the heat capacity as it contains dynamical input and no longer equals ({\it e.g.} at fixed volume) static energy fluctuations. First, the specific heat diverges when the Fermi energy lies between two kinetically defined energies. This divergence (or zero-temperature phase transition) is governed by the ratio between relaxation and dissipation times, as will be explained.
Secondly, there may appear a low-temperature regime of negative heat capacity. As will be seen, that can be understood from the occurrence of an anti-correlation between excess heat and configurational entropy.\\

The next section contains the details of the model. Section \ref{calo} explains the calorimetry and the method of calculation of the specific heat.
Explicit expressions of its thermal response are given in Section \ref{res}. Results include the possible divergence of the specific heat, giving a zero-temperature transition, and the possible negativity of the low-temperature specific heat.  We emphasize that in the context of interacting particle systems, the above low-temperature features can be expected to be more universal.  Calculations on the model are mostly postponed to the Appendix.

\section{Driven dynamics}\label{dr}
\begin{figure}[!t]
\centering
  \includegraphics[width=0.9\linewidth]{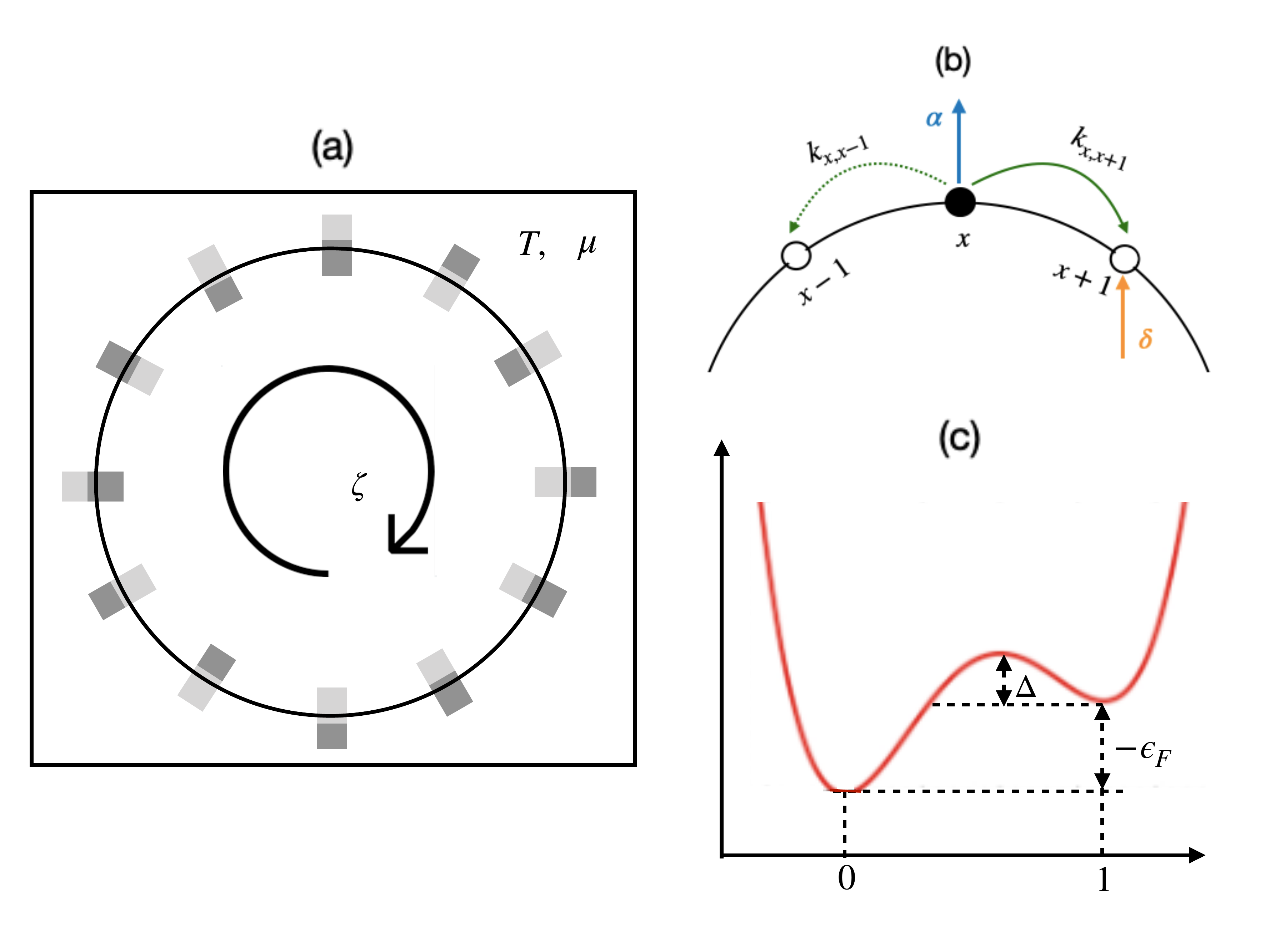}
 \caption{(a) Cartoon of a QD--array in an electron bath characterized by an ambient low temperature $T$ and chemical potential $\mu$.  The driving $\zeta$ breaks time-reversal invariance and a steady electronic current is maintained. (b) Scheme of transitions with their rates. There is a clockwise bias along the circuit.  (c) QD as a two-level system for negative Fermi energy. With kinetic barrier $\Delta$ and chemical potential $\mu=\epsilon_F$, the QD is charged at rate $\delta=e^{-\beta \Delta} e^{\beta \mu}$, and emptied at rate $\alpha=e^{-\beta \Delta}$.}
  \label{twolevel}
\end{figure}

Consider a ring with $N$ sites (stations or quantum dots (QD)), $x=1,2,\ldots, N$, each of which can be occupied by at most one particle. We write $\eta$ for the configuration and $\eta_x = 0$ or $1$, is the occupation at site $x$. See Fig.~\ref{twolevel}.\\
The dynamics is specified in terms of transition rates $W(\eta\rightarrow \eta')$ for the jump from configuration $\eta$ to $\eta'$.
There are two types of transitions: $\eta\rightarrow \eta^{x,x+1}$ where $\eta^{x,x+1}$ is the configuration defined by swapping the occupations at sites $x$ and $x+1$ in $\eta$, and $\eta\rightarrow\eta^x$ where $\eta^x$ is  obtained from $\eta$ by flipping the occupation at $x$.  The former transition indicates the hopping of particles (when $\eta_x\neq \eta_{x+1}$) and the latter stands for the onsite exit (when $\eta_x=1$) and entry (for $\eta_x=0$) of particles, both in  contact with a heat bath at inverse temperature $\beta$ and with chemical potential $\mu$.  QD have indeed the particularly useful possibility of
attaching current and voltage leads.\\
For the specific form of the rates, we follow the condition of local detailed balance, \cite{ldb}, and secondly, our modeling is semiclassical with incoherent coupling to an electron bath, taking a Fermi Golden Rule approximation.  More specifically, we put, with reference frequency $\nu$,
\begin{eqnarray}\label{hop1}
  W(\eta\rightarrow \eta^{x,x+1})&=& \frac{\nu}{1+e^{-\beta\zeta}}\;\textrm{ when } \eta_x =1, \eta_{x+1}=0\nonumber\\ &=&\frac{\nu}{1+e^{\beta\zeta}}\;\textrm{ when } \eta_x =0, \eta_{x+1}=1
 \end{eqnarray}
 for exchanging the occupation at sites $x\leftrightarrow x+1$,
where  $\zeta>0$ is the work to move one electron to a neighboring dot, say in  clockwise direction.  See Fig.\ref{twolevel}(b). 
The exclusion between the particles makes the dynamics fermionic.\\
%
A quantum dot also has a charging energy, required to add or remove a single electron from
the dot.  The loading and emptying of the quantum dot at $x$  is modeled as a birth and death process; see Fig.\ref{twolevel}(c).  At rate $\alpha=e^{-\beta \Delta}$ the particle is removed from the system, and with rate $\delta=\alpha\, e^{\beta\mu}$ the particle enters, which is summarized in the transition rate
  \begin{equation}
     W(\eta\rightarrow \eta^x) =
          \begin{cases}
            \tilde{\nu}\alpha & \textrm{ when } \eta_x =1 \\
           \tilde{\nu}\delta & \textrm{ when } \eta_x =0
          \end{cases}
  \end{equation}
  for the transition $\eta_x\rightarrow 1-\eta_x$.  Parameter $\tilde{\nu}$ sets the time scale of the birth/death process. We will also consider cases where $\Delta = \Delta(x,\eta)\geq 0$ is a configuration-dependent kinetic barrier.  For example, we may wish to facilitate the births and deaths at $x$  when at least one neighbor is occupied, and then put $\Delta(x,\eta) = 0$ when $\eta_{x+1} + \eta_{x-1} \geq 0$ (and equal to some positive number otherwise:  the kinetic barrier at $x$ is then only effective when its two neighbors are empty).  It adds a  local interaction, in the form of a density-dependent charging/emptying of the QD.
The chemical potential $\mu$ of the bath can be interpreted as an energy difference between two levels and is taken temperature-independent.  This allows abuse of notation, to take the chemical potential equal to the Fermi energy $\epsilon_F$, as we focus on low temperatures.\\  

The particles are not independent because of the exclusion and interact when the barriers are configuration-dependent barriers.  Taking the case of electrons, we still choose to ignore the possible influence of Coulomb repulsion.  {\it E.g.}, recently it was shown, at least for a pair of electrons trapped in a GaAs QD \cite{can} that Coulomb repulsion is not significant for thermal properties.\\

The Master Equation for the time-dependent probability  $P_t(\eta)$ reads 
\begin{equation}
\frac{\id P_t}{\id t}(\eta) =\sum_{\eta'}[W(\eta'\rightarrow \eta) P_t(\eta')-W(\eta\rightarrow\eta')P_t(\eta)]
\label{me}
\end{equation}
where again $\eta'$ is either $\eta^{x,x+1}$ or $\eta^x$ for some $x$, depending on the transition.
Putting Eq.~(\ref{me}) $ = 0$, the stationary probability distribution is
\begin{equation}\label{fdi}
P^s(\eta)\propto \,\exp\,[\beta\mu\,\sum_x\eta_x\,]
\end{equation}
  There is no detailed balance but $P^s(\eta)$  is a product of Fermi-Dirac distributions, independent of driving $\zeta$ and determined by the density $\rho = $Prob$[\eta_x=1] = (e^{-\mu\beta} + 1)^{-1}$. 
Nevertheless, as we will see and due to the driving, heat capacities become very different from their counterpart in equilibrium regimes.

\section{Calorimetry}\label{calo}
Calorimetry starts by identifying heat in the First Law of Thermodynamics.  Let $q(\eta,\eta')$ denote the heat released to the bath in the transition $\eta\rightarrow \eta'$.  Then, when in configuration $\eta$, the expected heat flux is
\begin{eqnarray}\label{dottqq}
\dot{q}(\eta) &=& \sum_{\eta'} W(\eta\rightarrow \eta')\, q(\eta,\eta')\\
&=&
\mu\tilde{\nu} e^{-\beta\Delta}\sum_x [e^{\mu\beta}(1-\eta_{x}) - \eta_x]+ \dot{w}(\eta) \nonumber
\end{eqnarray}
The first term in the second line is the rate of change of energy $E(\eta) = -\mu\sum_x \eta_x$ by loading or emptying the QD.  The second term involves the (Joule) heat produced by the work done by the driving force. More specifically, the net work done per unit time due to the driving is  
\[
\dot{w}(\eta) = \zeta \sum_x [W(\eta\rightarrow \eta^{x,x+1})\eta_x(1-\eta_{x+1}) -W(\eta\rightarrow \eta^{x,x+1})\eta_{x+1}(1-\eta_x)]
\]
Now substituting there the rates from \eqref{hop1}, one arrives at 
\begin{equation}\label{dottq}
\dot{q}(\eta) =
\mu\tilde{\nu} e^{-\beta\Delta}\sum_x [e^{\mu\beta}(1-\eta_{x}) - \eta_x]+ \nu\zeta \Gamma \sum_x \eta_x(1-\eta_{x+1})
\end{equation}
where we set $\Gamma = \sinh \beta\zeta\, (1+ \cosh\beta\zeta)^{-1}$.\\
Continuing from  \eqref{dottq}, the stationary heat flux (or steady dissipated power) is 
\begin{equation}\label{stath}
\dot{q}^s = \langle \dot{q}(\eta)\rangle^s = N\mu\tilde{\nu}\,[e^{(\mu-\Delta)\beta}\,(1-\rho) - e^{-\Delta\beta}\,\rho] +  N\zeta\nu\rho(1-\rho)\Gamma > 0
\end{equation}
Its positivity, which can be proven from convexity, is compatible with the Second Law: the entropy of the Fermi bath is never decreasing.\\

The time-accumulated difference between the instantaneous Eq.~(\ref{dottq}) and the stationary Eq.~(\ref{stath}) power is the excess heat, and can be quantified by the quasipotential 
\begin{equation}\label{qpi}
V(\eta) = \int_0^\infty \id t\,[e^{tL} \dot{q}\,(\eta) - \dot{q}^s]
\end{equation}
where $L$ is the backward generator of the Markov process: dual to \eqref{me},
\[ 
Lg(\eta) = \sum_{\eta'} W(\eta\rightarrow\eta')\,[g(\eta') -g(\eta)]
\]
for all functions $g$.  The convergence of the integral in Eq.~\ref{qpi} is exponentially fast for all fixed $N$ by the irreducibility of the Markov process.  Automatically, by taking stationary expectations, the quasipotential $V$ has a vanishing stationary expectation, $\langle V\rangle^s =0$.  Equivalently, by the identity
\[
-L\,\int_0^\infty \id t \;e^{tL}f = f - \langle f\rangle^s
\]
as applied to Eq.~(\ref{qpi}), the quasipotential $V$  verifies the Poisson equation: for all $\eta$,
\begin{equation}\label{lins}
LV(\eta)  
= \dot{q}^s -  \dot{q}(\eta)
\end{equation}
The left-hand side of Eq.~(\ref{lins}) can thus be read as the rate of change of the quasipotential.  
In equilibrium, $\zeta=0$,  $V(\eta) = E(\eta) - \langle E\rangle^s $ where $\langle E\rangle^s = \langle E\rangle^\textrm{eq}$ is the equilibrium energy.\\

The nonequilibrium heat capacity can be obtained from the quasipotential $V$.  After all, in \eqref{qpi} we see the quasipotential  as the accumulated excess in dissipated power.  That depends on the ambient temperature.  In general, the heat capacity is (with $k_B=1$),
\begin{equation}\label{hc}
    C(T) = \beta^2\,\biggl \langle \frac{\id V}{\id \beta} \biggr \rangle^s
\end{equation}
For details on the origin of this formula, which involves quasistatic changes in temperature, see \cite{eu,jir,jstat,ner,ner2}. Note that in general Eq.~(\ref{hc}) is not the derivative of the mean energy with respect to temperature. Only in equilibrium, when $V(\eta) = E(\eta) -\langle E\rangle^\textrm{eq}$, do we get
\[
C_\textrm{eq}(T) = -\beta^2\,\frac{\id \langle E\rangle^\textrm{eq}}{\id \beta}
\]

It may seem that in general, for steady nonequilibria, we need to solve the Poisson equation Eq.~(\ref{lins}) for obtaining the heat capacity Eq.~(\ref{hc}).  For small systems, like in \ref{r3}, to solve Eq.~(\ref{lins}) for $V$, we use numerically-assisted diagonalization. 
There is however another method, AC-calorimetry, introduced for nonequilibrium systems in \cite{jstat} and applied for active particles in \cite{active}, which is 
more stable and which we use for solving the problem for arbitrary ring sizes; see \ref{acv} for the details.
\begin{figure}[!t]
\centering
  \includegraphics[width=\linewidth]{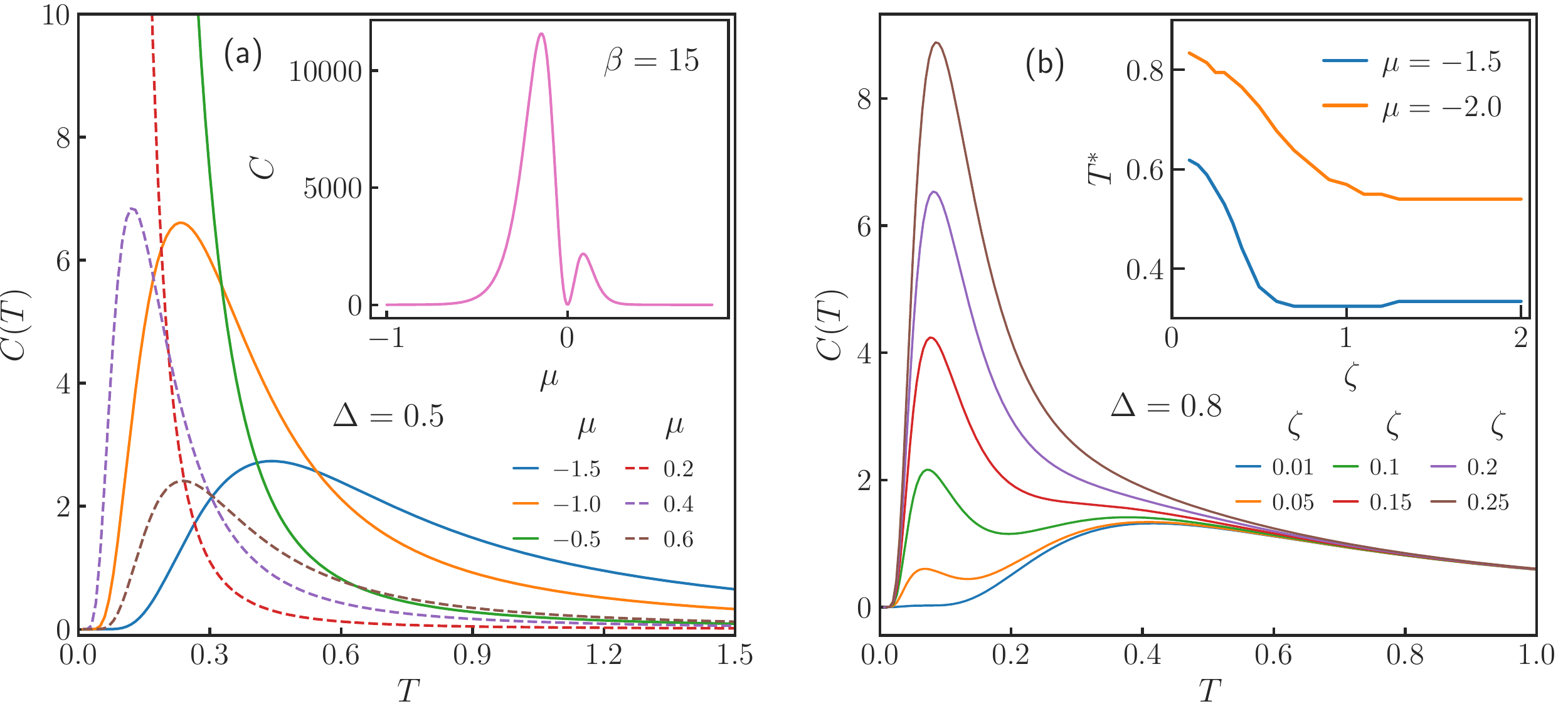}
  \caption{Heat capacity for $N=3$ (a)  for a uniform kinetic barrier $\Delta=0.5$ and $\zeta=1.0$\,. Inset: at  fixed $\beta=15$, $\Delta=0.5$ and $\zeta=1.0$\,. 
  (b) as function of driving $(\zeta)$ at  $\Delta=0.8$ and $\mu=-1.0$\,. Inset: Position of the peak temperature $(T^*)$   as function of $\zeta$ at $\Delta=0.8$\,.}
  \label{fig:C_8state_uniform}
\end{figure}

\section{Results}\label{res}
The exact computation in \ref{r3} yields the specific heat $C(T)/N = c(T)$,
\begin{eqnarray}\label{c3}
c(T) &=& \frac{\beta^2\mu^2 e^{-\beta\mu}}{(1 +e^{-\beta\mu})^2} \\ \nonumber
&-&\beta^2 \mu\zeta\frac{\nu}{\tilde{\nu}}\,\frac{ (e^{\beta(\Delta-3\mu)}-e^{\beta(\Delta-2\mu)})}{(1+ e^{-\beta\mu})^4}\tanh(\beta\zeta/2) \geq 0
\end{eqnarray}
Fig.~\ref{fig:C_8state_uniform}(a) shows its $\mu-$dependence. Note that the kinetic parameters $\nu,\tilde{\nu}, \Delta$, and the $\zeta-$dependence are  in the second (nonequilibrium) term. This information is visible at order $\zeta^2$ in the driving, most clearly at lower temperatures.  Note from Fig.~\ref{fig:C_8state_uniform}(b) the giant magnification of the peak for large $\zeta$, as compared to the Schottky peak for equilibrium.  The peak temperature (inset of Fig.~\ref{fig:C_8state_uniform}(b)) also saturates at a lower value of the temperature, as $\zeta$ grows.\\

From the inset of Fig.~\ref{fig:C_8state_uniform}(a), we see a diverging heat capacity which implies that the extended  Nernst postulate is violated.
 More precisely, the specific heat  $c(T\downarrow 0) \to \infty$ diverges with vanishing temperature for all values of the chemical potential $\mu$ with $-\Delta< \mu = \epsilon_F < \Delta/2$.  Otherwise,  $c(T\downarrow 0)\to 0$ as in the Third Law, \cite{ner, nernst2}.\\

The divergence is due to  dynamical localization, seen in the growth of relaxation times beyond the dissipative time scale; see \cite{ner,nernst2}.  For $\mu<0$, the empty state $\sum_x\eta_x=0$, and for $\mu>0$, the filled state $\sum_x\eta_x=N$, are overwhelmingly dominant at absolute zero.  The current carrying configurations ({\it e.g.} around half-filling) are suppressed which implies that the current vanishes as $e^{-\beta|\mu|}$ for $\beta\uparrow \infty$.  It sets the dissipation time $\tau_d\propto \nu^{-1}\, e^{|\mu|\beta}$.  Yet, the current-carrying configurations are separated from those ground states by the kinetic barrier: for $\mu<0$, it takes a time of the order $\tau \propto \tilde{\nu}^{-1}\,e^{\beta \Delta}$ to empty, and for $\mu >0$,  the relaxation time is $\tau \propto \tilde{\nu}^{-1}\,e^{\beta (\Delta-\mu)}$ to fill.  The Third Law (in the extension \cite{ner, nernst2}) holds when $\tau < \tau_d$.  At equality $\tau \simeq \tau_d$, we find $\mu=-\Delta$ or $\mu=\Delta/2$ and indeed, from Eq.~(\ref{c3}), at those two values a zero-temperature divergence occurs, unseen in equilibrium; see the inset of Fig.\ref{fig:C_8state_uniform}(a).\\

So far, the barrier $\Delta$ was always there, for no matter what configuration that would change its occupation number. 
We next install a configuration-dependent barrier, where
the reservoir at dot $x$ gets screened with a factor $e^{-\beta \Delta}$ if and only if both neighboring dots $x\pm 1$ are occupied. (The opposite case of a kinetic barrier at low density is analogous; see \ref{ddkb}).  Here we leave the strict context of quantum dots, and refer to the important physics of kinetic constraints as {\it e.g.} in glassy dynamics; see \cite{KinConSollich}.\\
For equilibrium, that would not change the specific heat.  Yet for nonequilibrium ($\zeta\neq 0$), the specific heat is different from Eq.~(\ref{c3}), see Appendix \ref{c3a},
and plotted in Fig.~\ref{fig:C_8state_nonuniform}(a). As $\beta\uparrow \infty$, 
$C\to -\infty$ for $-\Delta/4 < \mu < 0$, and $C\to \infty$  for $0 < \mu < \Delta/2$, when $\mu\, \zeta\,\nu\,\Delta \neq 0$.  The divergence for positive $\mu$ is much stronger as seen in the inset of Fig.~\ref{fig:C_8state_nonuniform}(a): the specific heat at $\mu<0$ is about 10 times larger than for $\mu>0$ at $\beta=15$.  
\begin{figure}[!t]
\centering
   \includegraphics[width=0.7\linewidth]{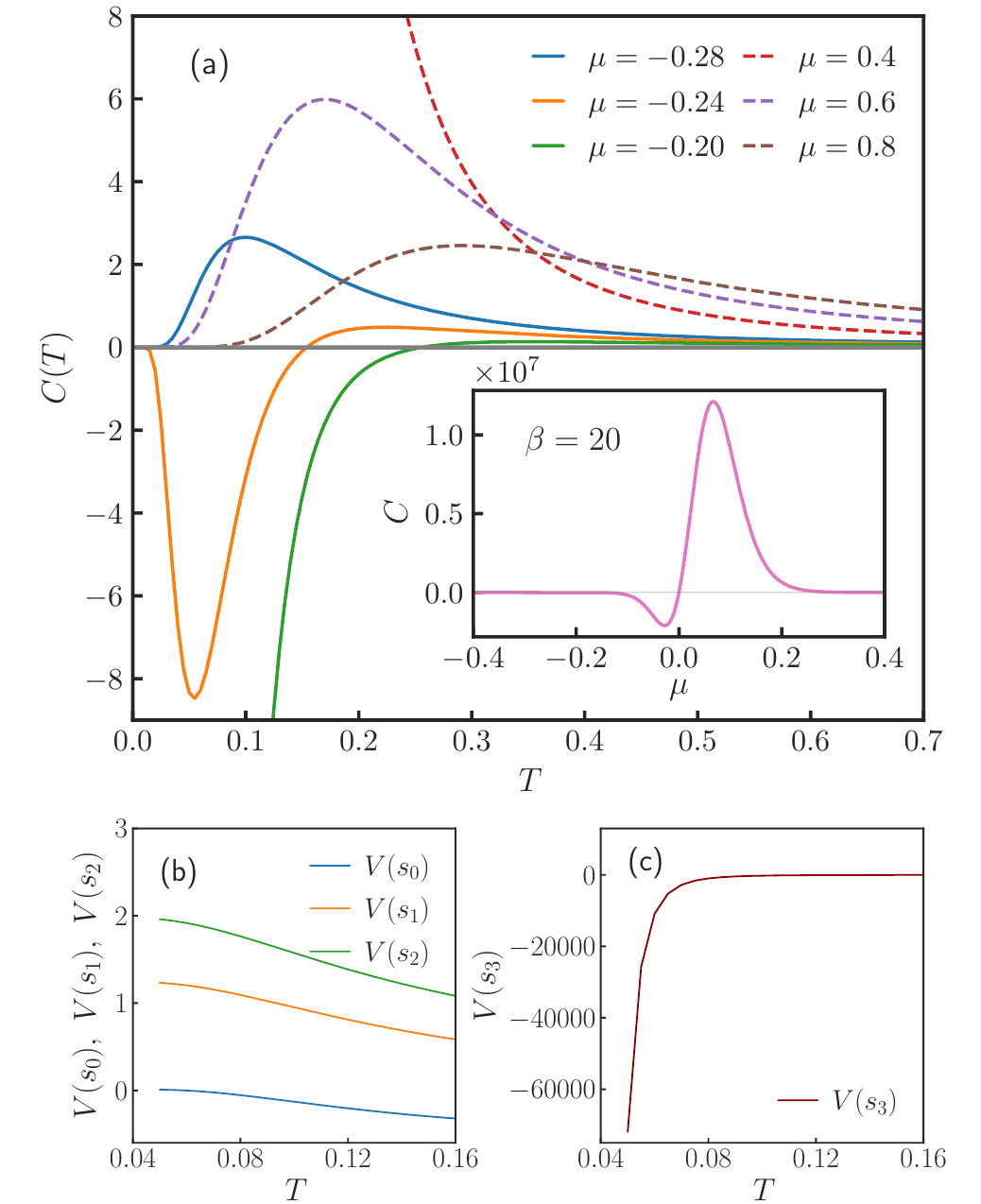} 
  \caption{Heat capacity in case of kinetic barrier at high density. (a) for different $\mu$ at $\zeta=1.0$ and $\Delta=0.8$.  Inset: as a function of $\mu$ at fixed  $\beta=20$, $\Delta=0.8$, and $\zeta=1.0$, showing the low-temperature divergence of the heat capacity. (b)-(c) Quasipotentials for $\Delta=0.8$, $\zeta=1.0$ and $\mu=-0.24$, corresponding to the orange curve in (a). }
  \label{fig:C_8state_nonuniform}
\end{figure}
That way again, the specific heat discloses the nature of a kinetic barrier, but an additional feature appears: the heat capacity gets negative.\\

To understand the negativity, we look at Eq.~(\ref{hc}), which we rewrite as the covariance
\begin{equation}\label{cova}
C(T) = \langle V
\,;\,\frac{\id \log P^s}{\id T}\rangle^s
\end{equation}
between a heat-related potential $V$ and the occupation-related $\log P^s$.
From Eq.~\ref{fdi}, and since $\langle V\rangle^s=0$, we have
\[
C(T) = -\frac{\mu\beta^2}{(1+ e^{\mu\beta})^N}\,\sum_\eta  V(\eta)\,\cal {N}(\eta)\,e^{\mu\beta\cal N(\eta)}
\]
where $\cal N(\eta)$ denotes the number of particles for configuration $\eta$.
Taking the system size $N=3$, let $s_i$ denote any configuration with $\cal N = i$ particles, $i=0,1,2,3$. Then,
\[
c(T) = -\frac{\mu\beta^2}{(1+ e^{\mu\beta})^3}\,[ e^{\mu\beta} V(s_1) +2\,e^{2\mu\beta}\,V(s_2) +e^{3\mu\beta}\, V(s_3)]
\]
To be specific, take $\mu<0$ and $\beta|\mu|\gg 1$.  As seen in Fig.~\ref{fig:C_8state_nonuniform}(b-c): $V(s_3)<0$ diverges at low temperatures, while $V(s_1)$ and $V(s_2)$ remain of order one:
\begin{equation}\label{v3}
V(s_3) \simeq -v\;e^{(-2\mu+ b)\,\beta}\,.
\end{equation}
For $\mu= -0.24$, the values are $v=0.99, b=0.08$.  Therefore, when $T\downarrow 0$,
\[
c(T) \simeq -\beta^2\mu\,e^{3\mu\beta}\, V(s_3) \simeq v\,\beta^2\mu\,e^{(b+\mu)\beta} < 0
\]
which explains the negativity.
The point is that $V(s_3) - V(s_2)$ 
$\downarrow -\infty$ for $\beta\uparrow \infty$, {\it i.e.}, the barrier (between $s_2$ and $s_3$) requires the heat bath to pump a lot of energy into the system to allow  energy decrease. That is the physical origin of a negative correlation in Eq.~\ref{cova} between a Clausius-related heat or entropy $\beta V$ and the configurational (Boltzmann-like) occupation $\log P^s$\,. In thermal equilibrium both are related to energy, yielding the variance of the energy in the equilibrium case of Eq.~\ref{cova}.\\

\section{Conclusions}
Applications of nonequilibrium physics abound at low temperatures. Yet, low-temperature kinetics is in general based on empirical information solely, \cite{aq}.  The present paper presents an exactly solvable many-body model, illustrating the main concepts and methods. We find indeed that heat capacities carry important kinetic information, {\it e.g.}, about barriers in the charging and discharging rates. In particular, a zero-temperature phase transition is observed where the heat capacity jumps from zero to infinity.  The transition as a function of the Fermi energy locates the energy barrier between quantum dots and leads and indicates where the relaxation time starts to exceed the dissipation time. Secondly, we get a detailed understanding of the occurrence of negative specific heat in terms of a negative covariance between the quasipotential (governing excess heat) and the temperature-dependence of the level occupations: higher energy states need to absorb heat to relax to lower energy.\\
 
While we expect that coherent coupling with quantum leads will modify the results, at least in some detail, the methodology and interest are of a wider scope in opening and exploring the thermal properties of a many-body nonequilibrium system, {\it e.g.}, for studies of  driven lattice gases, \cite{dlg1,dlg2,dlg3}.  From our proof of principle, we are optimistic that calorimetry will prove a valuable tool, next to others such as spectroscopy, to scan static and dynamical degrees of freedom that get excited in steady nonequilibria as a function of the ambient temperature.

\appendix

\section{Ring with three sites}\label{r3}
For simplicity, we start with the computation of the heat capacity for the smallest-size system, $N=3$ dots in Fig.~\ref{twolevel}(a).
There are 8 states for a possible total of 0, 1, 2 or 3 particles occupying the system.  By the symmetry of that small-sized system, we are allowed to work with those 4 classes and denote by $s_0, s_1, s_2$ and $s_3$ respective configurations from those classes.
\subsection{Configuration-independent kinetic barrier}
\begin{figure}[!h]
\centering
  \includegraphics[width=0.99\linewidth]{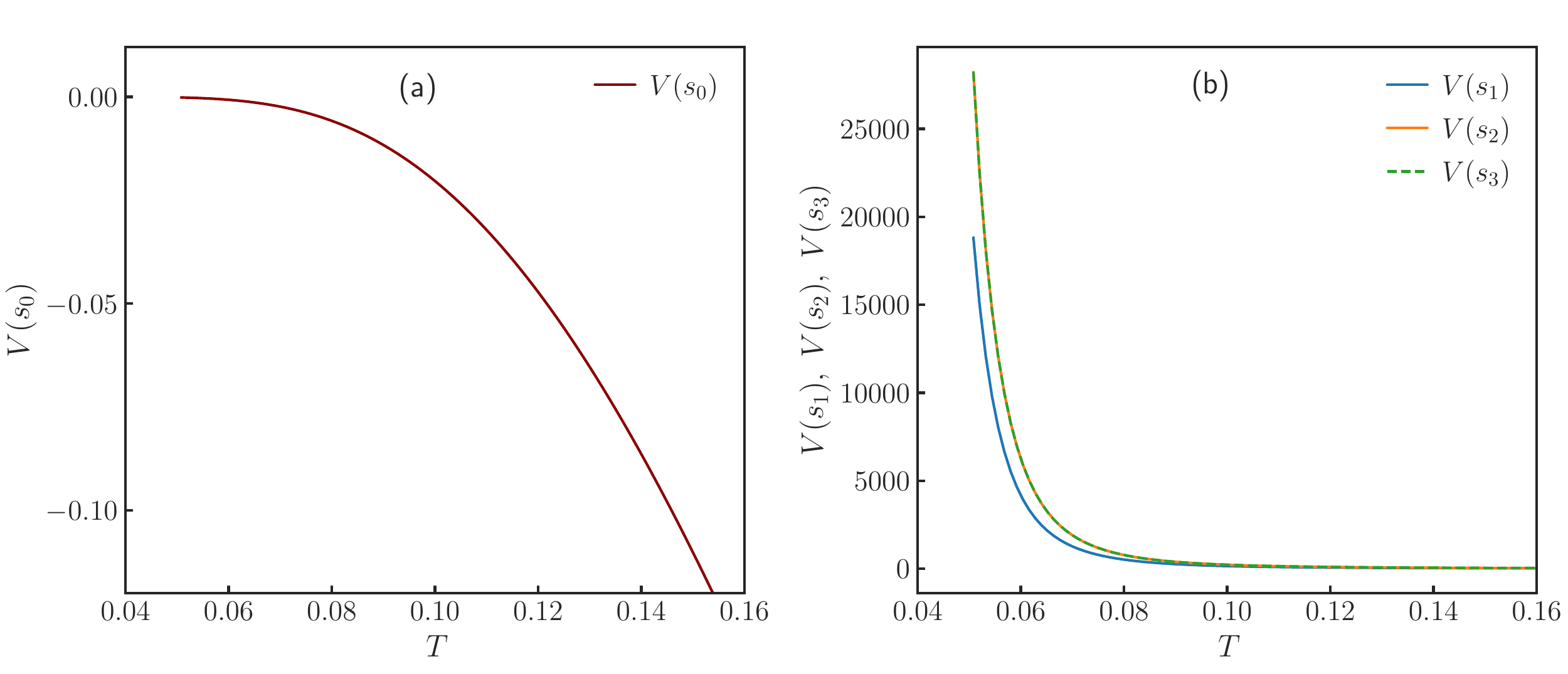}
  \caption{(a)-(b) Quasipotentials for uniform barrier height, $\Delta=0.5$, $\zeta=1.0$ and $\mu=-1.0$.  It corresponds to the orange curve in Fig.~2(a) in the main text.}
  \label{fig:quasi-potential-uniform}
\end{figure}

In the case that  the kinetic barrier holds for all birth/death transitions, $I[\eta_{x-1},\eta_{x+1}] \equiv 1$,
we have as expected heat fluxes (see Sec.~\ref{calo}), 
\begin{eqnarray*}
\dot{q}(s_0) = 3\mu\delta,  &\quad&
\dot{q}(s_1) = -\mu\alpha +2\mu \delta +\zeta\nu\Gamma\\
\dot{q}(s_2) = -2\mu\alpha + \mu\delta +\zeta\nu\Gamma, &\quad&
\dot{q}(s_3) = -3\mu\alpha\nonumber
\end{eqnarray*}
The stationary heat flux is 
\[
\dot{q}^s= \langle \dot{q}\rangle^s = 3\zeta\nu\Gamma\frac{\alpha\delta}{(\alpha+\delta)^2}
\]
The quasipotential $V$ in Eq.~(\ref{qpi}) satisfies Eq.~(\ref{lins}), which are coupled linear equations. The solution is plotted in Fig.~\ref{fig:quasi-potential-uniform}(a--b) for a particular choice of parameters.  Note  that $V(s_3)>V(s_2)>V(s_1)>V(s_0)$ are in the same order as their respective energies $E(s_i) = -\mu \,i$ for $\mu<0$.  This explains the positivity of the heat capacity, in contrast with the case of a non-uniform barrier as discussed around Eq.~(\ref{cova}).
The specific heat in Eq.~(\ref{c3}) is obtained from Eq.~(\ref{hc}).
The low- and high-density asymptotics $|\mu| \uparrow \infty$ of Eq.~(\ref{c3}) is given by
\begin{equation}\label{loh}
c_\textrm{low/high}(T) = \beta^2\mu^2 e^{-\beta|\mu|}
\end{equation}
and hence, in this case, the equilibrium (thermodynamic) contribution dominates. 

\subsection{Density-dependent kinetic barrier}\label{ddkb}
We now install a configuration-dependent barrier by making the reservoir screened with a factor $e^{-\beta \Delta}$ in the birth and death rates, if and only if both neighbors are occupied. The calculation proceeds along the same lines as before, with the expected heat fluxes 
\begin{eqnarray*}
\dot{q}(s_0) = 3\mu\delta,  &\quad&
\dot{q}(s_1) = -\mu\alpha +2\mu \delta +\zeta\nu\Gamma\\
\dot{q}(s_2) = -2\mu\alpha + \mu\delta\Omega +\zeta\nu\Gamma, &\quad&
\dot{q}(s_3) = -3\mu\alpha\Omega\nonumber
\end{eqnarray*}
where $\alpha=\tilde{\nu}$, $\delta=\alpha e^{\beta \mu}$, $\Omega=e^{-\beta \Delta}$\, and the stationary heat flux is 
\[
\dot{q}^s= \langle \dot{q}\rangle^s = 3\zeta\nu\Gamma\frac{\alpha\delta}{(\alpha+\delta)^2}\,.
\]
Quasipotentials are now given in Fig.~\ref{fig:C_8state_nonuniform}(b)-(c).\\  
The specific heat becomes 
\begin{equation}\label{c3a}
c(T) = \frac{ \beta^2\mu^2 e^{-\beta\mu}}{(1 +e^{-\beta\mu})^2} 
-\frac{\beta^2 \mu\zeta\nu}{\tilde{\nu}}\frac{ (e^{-5\beta\mu}+e^{-4\beta\mu}-e^{-3\beta\mu}-e^{\beta(\Delta-2\mu)})}{(1+ e^{-\beta\mu})^6}\tanh(\beta\zeta/2)
\end{equation}
and plotted in Fig.~\ref{fig:C_8state_nonuniform}(a)  for different values of $\mu$.\\

\begin{figure}[!t]
\centering
  \includegraphics[width=0.7\linewidth]{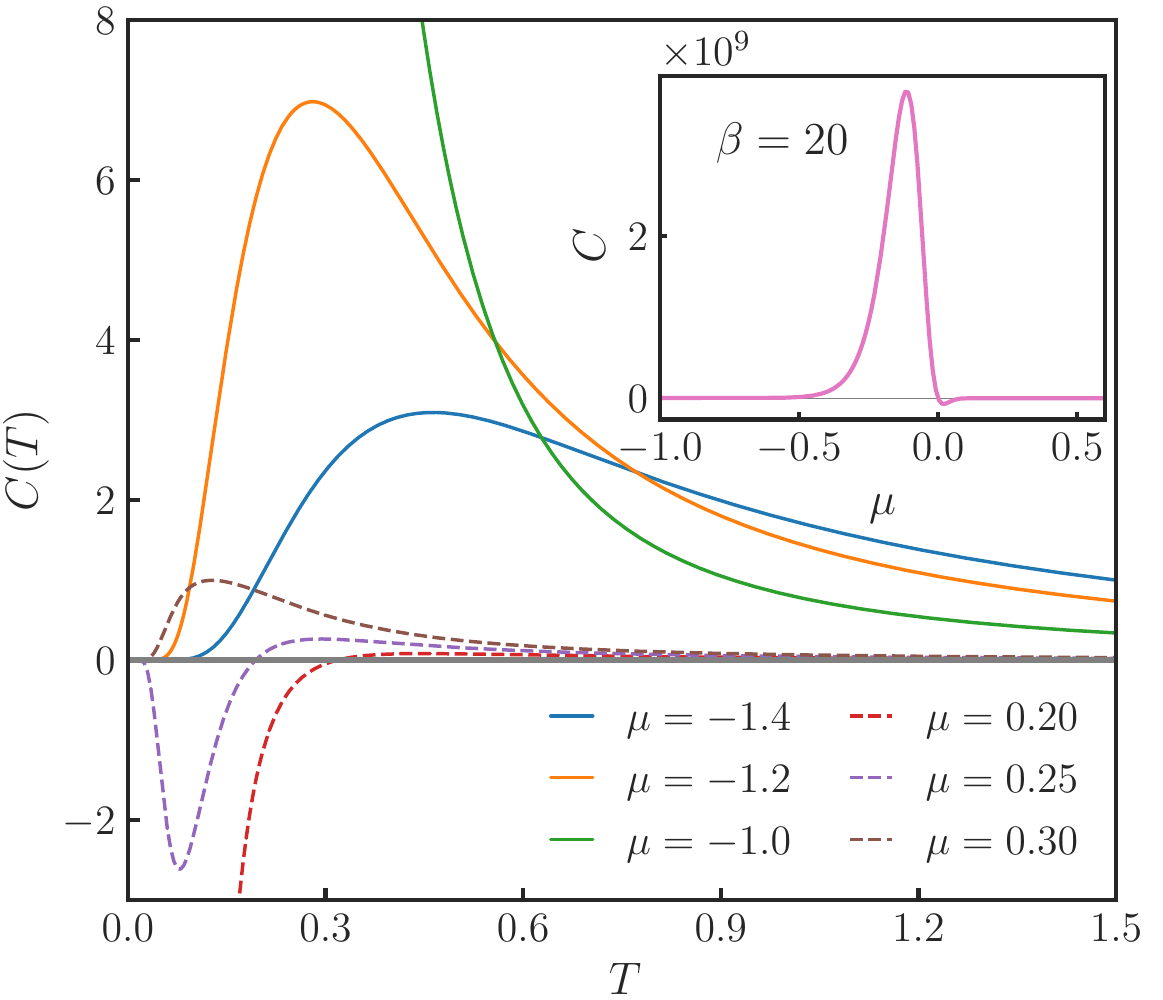}  
  \caption{Heat capacity for $N=3$ with density-dependent kinetic barrier. $C(T)$ for different $\mu$ at fixed $\Delta=1.0$ and $\zeta=1.0$. In this case, birth and death rates are obstructed when both the neighboring sites are empty. Inset: Heat capacity as a function of $\mu$ at fixed inverse temperature $\beta=20$, $\Delta=1.0$ and $\zeta=1.0$ }
  \label{fig:C_empty}
\end{figure}

The same can be done for a kinetic barrier obstructing birth and death at $x$ when both neighbors are unoccupied,  In that case, expected heat fluxes are 
\begin{eqnarray*}
\dot{q}(s_0) = 3\mu\delta\Omega,  &\quad&
\dot{q}(s_1) = -\mu\alpha\Omega +2\mu \delta +\zeta\nu\Gamma\\
\dot{q}(s_2) = -2\mu\alpha + \mu\delta +\zeta\nu\Gamma, &\quad&
\dot{q}(s_3) = -3\mu\alpha\nonumber
\end{eqnarray*}
where $\alpha=\tilde{\nu}$, $\delta=\alpha e^{\beta \mu}$, $\Omega=e^{-\beta \Delta}$\, and the stationary heat flux remains the same as before. 
After an exact calculation, the specific heat  now becomes
\begin{equation}\label{c3b}
c(T)=\frac{\beta^2\mu^2 e^{-\beta\mu}}{(1 +e^{-\beta\mu})^2}
-\frac{\beta^2 \mu\zeta\nu}{\tilde{\nu}}\frac{ (e^{\beta(\Delta-5\mu)}+e^{-4\beta\mu}-e^{-3\beta\mu}-e^{-2\beta\mu)})}{(1+ e^{-\beta\mu})^6}\tanh(\beta\zeta/2)
\end{equation}
Again, we see the possibility of negative heat capacities;  see Fig.~\ref{fig:C_empty}.  Concerning the Nernst postulate, for $\beta\uparrow \infty$, $C\to \infty$ for $-\Delta < \mu < 0$ and $C\to -\infty$ for $0 < \mu < \Delta/5$ if $\mu\, \zeta\,\nu\,\Delta \neq 0$.   Now, the divergence is much stronger at negative values of the chemical potential;  see the inset of Fig.~\ref{fig:C_empty}.  

Note that the low-density regime where the exclusion principle plays a minor role, does not differ from \ref{loh}. \\

The same method can be followed for every ring size $N$ but it becomes of course computationally challenging to solve the $2^N$ equations Eq.~(\ref{lins}) for the quasipotential $V$. 
We, therefore, introduce another method that appears more flexible, is applicable for all $N$, and, so we believe, gives
the most promising experimental setup.  We refer to \cite{jstat,active,sul,ND} for the background of the AC-calorimetric method that we next apply.

\section{AC-calorimetric method}\label{acv}
For simplicity, we put $\tilde{\nu}=1$ and work with a uniform kinetic barrier.\\
 We consider a sinusoidal modulation in temperature  $T_t = T+\tilde{\epsilon} \sin\omega t$\, at frequency $\omega$ for small amplitude $\tilde{\epsilon}$. As a consequence, the  birth and death rates become $\alpha_t = \alpha\,(1+ \epsilon_a\sin \omega t), \delta_t = \delta\,(1+ \epsilon_d\sin \omega t)$ with 
 \begin{equation}
    \epsilon_a = \frac{ \tilde{\epsilon}\Delta}{T^2},\,\,\,\epsilon_d = \frac{ \tilde{\epsilon}(\Delta-\mu)}{T^2},\quad \alpha=e^{-\Delta/T}, \;\;\delta = \alpha e^{\mu/T}
\end{equation}
Now there is a time-dependent heat flux to the bath, 
\begin{equation}\label{qt}
    \dot{Q}(\rho_t) =\zeta\nu\rho_t(1 -\rho_t) \Gamma(\beta_t\zeta) - \mu \left[ \rho_t(\alpha_t +\delta_t) -\delta_t \right]
\end{equation}
obtained from the expression of the heat flux in Eq.~(\ref{dottq}) 
but averaged with the time-dependent probability distribution $\rho_t$ solving
\begin{equation}\label{td}
    \frac{\id\rho_t}{\id t}=-(\alpha_t + \delta_t)\rho_t + \delta_t\,.
\end{equation}
Working in the complex domain of temperatures, a rather long but straightforward  calculation gives the large--time and small-frequency limit, 
\begin{equation}
    \rho(t)=\frac{\delta}{(\alpha +\delta)} -\tilde{\epsilon}\frac{\delta}{(\alpha+\delta)^3}\biggl(\frac{\alpha\delta}{T^2} \biggr)\left[\omega +i(\alpha+\delta) \right]e^{i\omega t}
\end{equation}
to linear order in $\tilde{\epsilon}$.\\
Finally, from general results \cite{jstat,active,ND}, the out-of-phase component in $\dot{Q}(\rho_t)$  gives the heat capacity.  In these limits indeed, the heat flux \ref{qt} verifies
\begin{equation}\label{oof}
    -\dot{Q}(t) = -\dot{q}^s + \tilde{\epsilon} \,[B(T)\, \sin(\omega t) + C(T)\,\omega\, \cos(\omega t)]
\end{equation}
That can be calculated exactly, and we  find that the specific heat coincides with the expression Eq.~(\ref{c3});  
thus it gives the correct result for all $N$ and all conclusions (and Fig.~\ref{fig:C_8state_uniform}) remain unaltered.\\

\noindent{\bf Acknowledgment:}  For many clarifying discussions we are indebted to Faezeh Khodabandehlou and to Karel Neto\v{c}n\'{y}.

\end{document}